\documentclass{article}
\usepackage{times}
\usepackage{graphicx}
\usepackage{latexsym}
\usepackage{url}
\usepackage{amsmath}
\usepackage{graphicx}
\usepackage{tabularx}
\usepackage{array}

\begin{document}
\title{Trust dynamics and user attitudes on recommendation errors: preliminary results
}

\author{David A. Pelta , Jos\'e L. Verdegay, Mar\'ia T. Lamata, Carlos Cruz Corona
	\thanks{Authors are with the Department of Computer Science and A.I, Models of Decision and Optimization Research Group, Universidad de Granada, Spain. e-mail: \{dpelta, verdegay\}@ugr.es, \{mtl,carloscruz\}@decsai.ugr.es}}

\maketitle

\begin{abstract}
	
	Artificial Intelligence based systems may be used as digital nudging techniques that can steer or coerce users to make decisions not always aligned with their true interests.
	When such systems properly address the issues of Fairness, Accountability, Transparency, and Ethics, then the trust of the user in the system would just depend on the system's output. 
	
	The aim of this paper is to propose a model for exploring how good and bad recommendations affect the overall trust in an idealized recommender system that issues recommendations over a resource with limited capacity. The impact of different users attitudes on trust dynamics is also considered.

	Using simulations, we ran a large set of experiments that allowed to observe that: 1) under certain circumstances, all the users ended  accepting the recommendations; and 2) the user attitude (controlled by a single parameter balancing the gain/loss of trust after a good/bad recommendation) has a great impact in the trust dynamics. 
	
\end{abstract}

\section{Introduction}

Decision making is an ubiquitous task that is not done in vacuum. Our decisions are constrained by our own preferences, by our social network, by the context, by the environment and so on. Moreover, we are surrounded by little nudges: indirect suggestions that are generated by some external agent to influence the decision making process of groups or individuals \cite{nudge2008}.

An example are the so called ``Dark Patterns'' defined as ``user interface design choices that benefit an online service by coercing, steering, or deceiving users into making unintended and potentially harmful decisions'' \cite{Mathur2019DarkPatterns}. In such paper,   
authors make recommendations to study, mitigate, and minimize the use of these patterns.
These and other nudging techniques are gaining attention in the last years \cite{WeinmannSchneiderBrocke2015, SchneiderWeinmannBrocke2018}

But these nudges can be issued by more sophisticated systems which through the use of proper data collection, modeling and learning are able to exploit our ``history'' and preferences trying to induce us taking some decisions. In what follows, we will refer to these systems as ADM: Automated Decision Making systems .

Nowadays, there is an increasing concern on how such ADM are designed and deployed and several countries, research centers and institutions are devoting efforts on how to best address this concern. 
As relevant examples we can cite 

As relevant examples we can cite a white paper (written by Informatics
Europe and the ACM Europe Policy Committee), presenting specific
recommendations from the European technical and scientific community about how policy makers, legislators, and concerned individuals might best respond to the rapid growth of ADM \cite{LarusHankinCarsonEtAl2018}.
We can also mention the European approach to Artificial Intelligence, 
where there is a High-Level Expert Group on Artificial Intelligence (AI-HLEG)\cite{HLE2018} that recently (June 2019) presented their Policy and Investment Recommendations for Trustworthy AI \cite{Policy2019} during the first European AI Alliance Assembly\footnote{\url{https://ec.europa.eu/digital-single-market/en/news/first-european-ai-alliance-assembly}}.
Besides these initiatives, the interested reader may find in \cite{Dutton2019}, a summary of 26 strategies for A.I. from different countries worldwide.

Overall, many of the principles, recommendations and guidelines can be summarized in four key issues: Fairness, Accountability, Transparency, and Ethics. 

If we focus on the people using those ADM, the concept of ``trust'' emerges as one of the most relevant ones. 
The European Policy mentioned before makes clear that trust is a prerequisite to ensure a human-centric approach to AI and identify seven key requirements that AI applications should respect to be considered trustworthy.

Why a user should trust the output (a decision, a recommendation) of such ADM? how good/bad decisions/recommendations affect the level of trust on the behavior of the system? How aspects like data collection, privacy management, strategic manipulation, nudging and so on affect trust? These are all relevant questions.

In this context, one may argue that as the ADM start to behave following the key issues mentioned above, the trust of the people in the system will be only affected by the system's output, as other aspects (like privacy management) will be properly managed by some external certification authority.

It is well known that defining trust is far from trivial as different disciplines define it differently. Here, and considering the so called recommender systems \cite{Jugovac2017,VILLEGAS2018173,LU20121} as a particular case of an ADM, we 
define trust as the willingness of the user to accept a recommendation based on a subjective belief that the recommender tool will exhibit reliable behavior to maximize the user’s interest under uncertainty of a given situation, based on past experiences with the tool. Our definition resembles the one presented in \cite{Cho2015}. 

Thus, the aim of this paper is to study the dynamics of trust on an multi-user scenario with an ideal recommender system.
We depart from a fair and unbiased idealized recommendation tool which issues binary suggestions \textsl{on using or not a resource with limited capacity}. Typical examples are take/do not take a given route, go/do not go to a restaurant or a bar, and so on. 
The use of a limited capacity resource forces the use of different recommendations even for users with the same profile. 
Consider, for example, route navigation apps. If all the drivers are recommended an alternative route to avoid a traffic jam ahead, then, the alternative route will be also congested within a short period of time and can generate disturbances in the neighborhoods where the traffic was diverted. So, it becomes clear that not all the users should receive the same recommendation (this is an aspect related with ``fairness'').

In our model, users simultaneously receive a recommendation on going or not going to a bar. The users decide at the same time whether they will go to the bar or not. As the bar has a limited capacity, it's no fun to go there if it is too crowded.
Every user has a level of trust on the recommender that is increased/decreased if the recommendation was good or not. 
The amount of increase/decrease in the level of trust is the key to model different user attitudes\footnote{
	This model resembles the ``El Farol'' bar problem \cite{ElFarol94}, a typical example of the so called Minority Games \cite{MinorityGames2013}. Here, we eliminate some assumptions like the existence of payoffs (in terms of game theory), a history of bar attendances and the use of several prediction strategies by the users. In turn, all the users employ the same trust based decision rule, while the recommender system (not present in the original problem) uses a very simple recommendation strategy.}.

We will explore how good/bad recommendations may affect the trust in the recommender, considering  different users attitude towards recommendation errors: tolerant, neutral or intolerant (in a continuum and not as discrete categories). 
In situations of repeated interactions, we will analyze how the overall trust in the recommender evolves and how the users attitude significantly affects the results. 

Using simulations, experiments will be done and conclusions will be outlined.



Consequently, the paper is structured as follows. 
In Section \ref{model} the components and the inner working of proposed model is described. 
Then, in Section \ref{experiments} the main experiments and results are described and analyzed. They are related with a) the evolution of  trust in the recommendations, and b) the influence of the user attitude on trust dynamics. 
Finally, Section \ref{conclus} is devoted to conclusions and further work.


\section{Model Description\label{model}}

The proposed model is based on three components: a resource, the users and the recommender.
These components are described below and then, the interactions among them is presented. 

\begin{figure*}
	\centering
	\begin{tabular}{cc}	
		\includegraphics[width=0.8\linewidth]{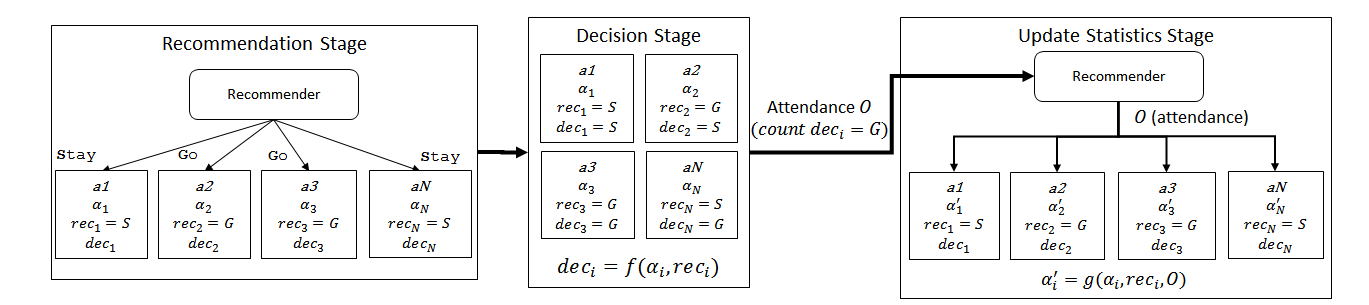} \\
		\includegraphics[width=0.8\linewidth]{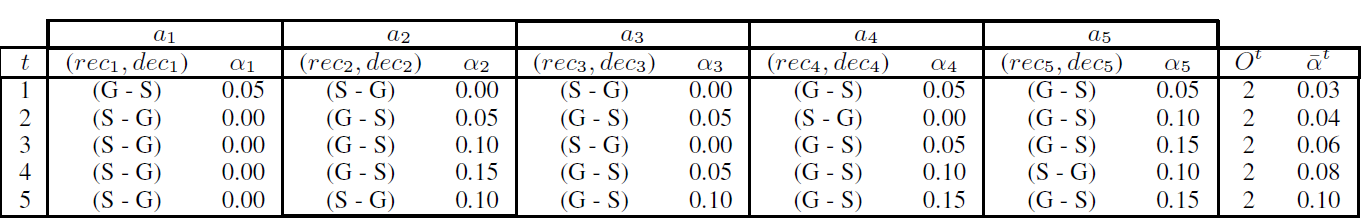} \\
	\end{tabular}	
	\caption{Top: Graphical representation of the model. Bottom: Example showing five iterations of the model with $N=5$ users and $L=3$. Recommendation, decision and level of trust for every user are shown. Also the attendance $O^t$ and the average trust $\bar{\alpha}^t$ are displayed in the last two columns.}
	\label{fig:esquema}
\end{figure*} 

\subsection*{a) The Resource}
We depart from a resource with a limited capacity (let's suppose a bar) $C$ and a ``comfort level'' $L$ (as in the El Farol problem \cite{ElFarol94}) which is the maximum number of users that makes the place not crowded.
We assume that a number $N > L$ of users exist (not all the users can simultaneously go to the bar).

The attendance $O^t$ is the number of users that decided to go to the bar at time $t$.

\subsection*{b) Users}
We have a set of homogeneous users $A = \{a_1,a_2, \ldots, a_N\}$, where 
every $a_i$ has:
\begin{itemize}
	\item $\alpha_i^t \in [0,1]$: level of trust on the recommendation at time $t$. 
	\item $recom_i^t, dec_i^t \in \{GO, STAY\}$: recommendation received and the decision taken at time $t$, respectively.
\end{itemize}

\noindent
\textbf{Decision Rule}: the user will accept the recommendation ($dec_i^t = recom_i^t$) with a value proportional to $\alpha_i^t$.
If the user rejects the recommendation, it will do the other action.


Notice that when $\alpha_i^t = 1$, the user will always accept the recommendation, but when $\alpha_i^t = 0$, it will do the opposite of the recommendation.\\

\noindent
\textbf{Trust Revision Protocol:}
every user has a protocol to modify its level of trust $\alpha_i^{t+1}$ in the recommender in terms of the last recommendation received ( $recom_i^t$) and the last attendance to the bar ($O^t$).

The users have two parameters $\beta, \gamma$ called the positive and negative feedback (or learning factor) respectively. 
In this initial setting, all users are considered homogeneous (they share the same decision rule and the trust revision protocol).\\

\noindent
The trust will increase, making $\alpha_i^{t+1} = \alpha_i^t + \beta$ if the recommendation was ``good''. This will happen either when:

\begin{enumerate}
	\item  the tool recommended to GO and the bar was not crowded ($recom_i^t = GO$ and $ O^t \leq L$), or 
	\item the tool recommended to STAY and the bar was crowded ($recom_i^t = STAY$ and $ O^t > L$)
\end{enumerate}

\noindent
In turn, the trust will decrease, making $\alpha_i^{t+1} = \alpha_i^t - \gamma$ if the recommendation was ``bad''. Either when:

\begin{enumerate}
	\item the tool recommended to GO and the bar was crowded ($recom_i^t = GO$ and $ O^t > L$), or  
	\item the tool recommended to STAY and the bar was not crowded ($recom_i^t = STAY$ and $ O^t \leq L$)
\end{enumerate} 

The use of different values for $\beta, \gamma$ has two reasons: 1) trust can be gained or lost at different rates, 2) the relation between both parameters allows to represent different user attitudes, leading to: 

\begin{itemize}
	\item \textit{Neutral User}: $\gamma = \beta$, the same feedback is added/substracted to the current level of trust.
	\item \textit{Tolerant User}: $\gamma < \beta$, the loose of trust occurs slower than trust gain: which means that the agent is tolerant to recommendation errors.
	\item \textit{Intolerant User}:  $\gamma > \beta$: the agent penalizes the recommendation errors. For example, when $\gamma = 2 \times \beta$ then an error in the recommendation has two times more impact in the level of trust than a good recommendation. In other words, after a bad recommendation, the user will need two good ones to recover the original level of trust. 
\end{itemize}

\subsection*{c) The Recommender}

The recommender knows the set of users but does not have access to their internal levels of trust (the $\alpha_i^t$ values). 
Just the last decision taken by every user $a_i$ (the value $dec_i^t$) is available to the recommender.

Given that the users are homogeneous, a profile based recommendation would not be possible (remember that the bar has a limited capacity). So, as a starting point, the recommender uses a very simple rule for the assignment of recommendations:

\begin{itemize}
	\item randomly select a set $G$ of $L$ users.
	\item every user $a_i \in G$ receives a $GO$ recommendation
	\item every user $a_i \notin G$ receives a $STAY$ recommendation
\end{itemize}

This recommender would be an ideal one from the user perspective: it has no room for manipulation, its behavior is clearly unbiased, it does not have access to the user's private information, it does not store any users' historical data and so on.

\subsection*{Working scheme\label{scheme}}
The elements of the model are depicted in Fig. \ref{fig:esquema} (top).


\noindent
At every time step $t$, there are three stages. 

\begin{enumerate}
	\item \textit{Recommendation Stage}: the recommender sends a recommendation $rec_i^t$ to every user $a_i$.
	\item \textit{Decision Stage}: using the decision rule described previously, every user takes a decision $dec_i^t$. As expected, the recommendation can be followed or not. 
	\item \textit{Update Statistics Stage}: taking into account the users decisions, some measures are calculated (see below) and informed to the users. Then they adapt their levels of trust using the revision protocol described previously.
\end{enumerate}

At every time step $t$ the following measures are calculated.

\begin{itemize}
	\item Attendance $O^t$: number of users that decided to go to the bar.
	\item Average trust on the recommendations: 
	 \begin{equation*}
	  	\bar{\alpha}^t = \frac{1}{N}\sum_{i=1}^N \alpha_i^t
	 \end{equation*}
\end{itemize}

Please note that the value $\bar{\alpha}^t$ is just informative and does not affect neither the decision of the users, nor the way recommendations are issued.

Let's suppose $N=5, L=3, \beta = \gamma = 0.05$ (neutral users). An example with five iterations is displayed in Fig. \ref{fig:esquema} (bottom).
For each user, three values are shown: $(rec_i, dec_i) \:\: \alpha_i$. We use the value `G' in $rec_i^t$ or $dec_i^t$ to denote a GO recommendation or decision while `S' states a STAY one.
Then, the attendance $O^t$ and the average trust $\bar{\alpha}^t$ appear.

Consider user $a_1$ when $t=1$ (first row). The recommendation was GO but the user decided to STAY. As the bar was not crowded ($O^t = 2$), the recommendation was good so the level of trust of $a_1$ is increased. 
In turn, consider user $a_3$. 
The recommendation was STAY but the user decided to GO. As the bar was not crowded ($O^t = 2$), the recommendation was bad. Trust should be decreased but as it could not be lower than zero, it stays in the minimum possible value.

We can also consider the dynamic behavior of every user. If we focus on user $a_5$, we observe that for $t=1,2,3$, it received a GO recommendation that the user does not follow. In those time steps, the bar was not crowded, so the level of trust of the user was increased. 
When $t=4$, the recommendation was STAY but it decided to GO. The bar was not crowded, so the trust was reduced. 
A similar analysis can be done for the rest of users. 

In this example, the average trust increased in every iteration.

It is important to remark that as the recommender has no access to the (private)  level of trust of the users, it can not
broadcast any sort of average trust to them. Neither is possible the communication among the users. Both aspects, although important, would add additional features to the model that may affect the analysis of the  trust dynamics.

\section{Experiments\label{experiments}}
Two experiments are conducted. The first one is aimed at understanding trust dynamics (how the average trust change with the time), while the second one focuses on how trust dynamics is affected by the user attitudes with respect to recommendation errors.

\subsection{The evolution of the average level of trust \label{exp1}}

This experiment is aimed to understand how the trust changes with the time. Some preliminary experiments showed us that the average level of trust converge to 1, so here we pose the following questions:
\begin{enumerate}
	\item \textit{Is there any $t^*$ which makes the individual values $\alpha_i^{t^*}, \: \forall \: i \in [1 \ldots N]$ converge?.}
	\item  \textit{In such a case, the convergence value is the same for all the users?. }
	\item \textit{Does the number of users has any implications on the results?}
\end{enumerate}

For different values of $N$, $L = 0.6 \times N$, and $\gamma = \beta = 0.05$ and a maximum of 250 iterations, we run 100 independent repetitions of the simulation. 

Results are shown in Table \ref{tab:time2eq}, where for each value of $N$, the number of repetitions that converged ($runs$), out of 100, and the average number of iterations done to converge ($I2C$) are displayed.

The first element to highlight is that all the repetitions converged, and the average level of trust reached the value 1. In other words, always $\exists \:t^*  \:|\: \alpha_i^{t^*} = 1, \:\: \forall i \in [1 \ldots N]$, where $t^*$ is the time (or iterations) to convergence.
This is extremely relevant because when such situation occurs, all the users will accept the recommendation, which means that (from the point of view of the recommender and the resource usage) the problem became an assignment problem instead of a recommendation one.

\begin{figure}
	\centering
	\includegraphics[width=0.9\columnwidth]{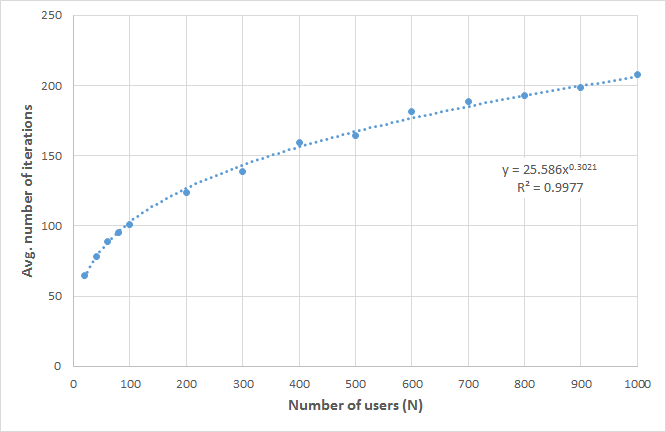}
	\caption{Average iterations to converge in terms of the number of users}
	\label{fig:NvsIters}
\end{figure}

Another point to analyze is the relation between the number of users $N$ and the average number of iterations to converge $I2C$. The plot in Fig. \ref{fig:NvsIters} shows this relation, which perfectly adjusts to a power law with $y = 25.586 x^{0.3021},\: R^2 = 0.9977$.

\begin{table}
	\centering	
	\begin{tabular}{r r c r r r}
		\hline	
N    &	L  & $runs$ & $I2C$ & stdDev\\
		\hline
20   & 12  & 100 & 64.61 & 8.08 \\
40   & 24  & 100 & 78.29 & 10.04 \\
60   & 36  & 100 & 88.96 & 11.89 \\
80   & 48  & 100 & 95.00 & 12.68 \\
100  & 60  & 100 & 101.03 & 13.60 \\
200  & 120 & 100 & 123.85 & 19.50 \\
300  & 180 & 100 & 138.81 & 19.05 \\
400  & 240 & 100 & 159.41 & 20.07 \\
500  & 300 & 100 & 164.58 & 26.72 \\
600  & 360 & 100 & 181.74 & 30.34 \\
700  & 420 & 100 & 188.29 & 32.58 \\
800  & 480 & 100 & 192.70 & 27.86 \\
900  & 540 & 100 & 198.71 & 33.95 \\
1000 & 600 & 100 & 208.01 & 39.90 \\
		\hline
	\end{tabular}
	\caption{Results from 100 repetitions for different number of users. All the repetitions converged. The average number of iterations to convergence ($I2C$) and the corresponding standard deviation are shown in the last two columns.
		\label{tab:time2eq}}
\end{table}

\subsection{On the influence of users' attitude \label{exp2}}

Now, in this experiment, we explore how the trust dynamics change in terms of the users attitude. 
The question we posed here is:\textit{ Does the user attitude (the relation between $\gamma$ and $\beta$) has any impact in the time to convergence?}.

Recall that a good recommendation makes $\alpha_i^{t+1} = \alpha_i^t + \beta$; otherwise trust changes as $\alpha_i^{t+1} = \alpha_i^t - \gamma$.

We fix $N=100, L=60$. We keep $\beta = 0.05$ and we define $\gamma$ as $\phi \times \beta$ with $\phi = \{0.5, 0.6, \ldots 1.8, 1.9, 2.0\}$

The $\phi$ value allows to model the user attitude as a continuum between tolerant to intolerant attitude.
When $\phi=1$, then a neutral user is modeled while $\phi < 1$ allows to model a tolerant one. 
Finally, when $\phi > 1$ an intolerant user is obtained.
For each value of $\phi$ we run 100 repetitions of the simulation, each one with a maximum of 5000 iterations. 

The results are shown in Table \ref{tab:influNegFeedback2}. 
Focusing first in the left part of the Table, two different behaviors are clearly observed. 
The first one is when $\phi \leq 1.4$, where all the runs converged.
In these cases, a clear exponential relation is observed between $\phi$ ($\gamma$) and the time to converge. 
When the negative feedback is lower than the positive one $\gamma < \beta$ (i.e. $\phi < 1$) the time to converge is shorter than when $\gamma = \beta$ ($\phi=1$). These would be the behavior of ``tolerant'' users that forgive the recommendation errors. 
When $1 < \phi \leq 1.4$ we are in the presence of users that are less tolerant  to recommendation errors. The higher the $\gamma$ (the negative feedback), the harder the convergence. 

When $\phi = 1.5$ an important change in the behavior of the model appeared. Just 60 \% of the runs converged while such percentage reduced to just 3\% when $\phi = 1.6$ . Moreover, when $\phi > 1.6$, the simulations did not converge within the iterations limits posed. 

\begin{figure}
	\centering
	\includegraphics[clip=true,width=0.77\linewidth]{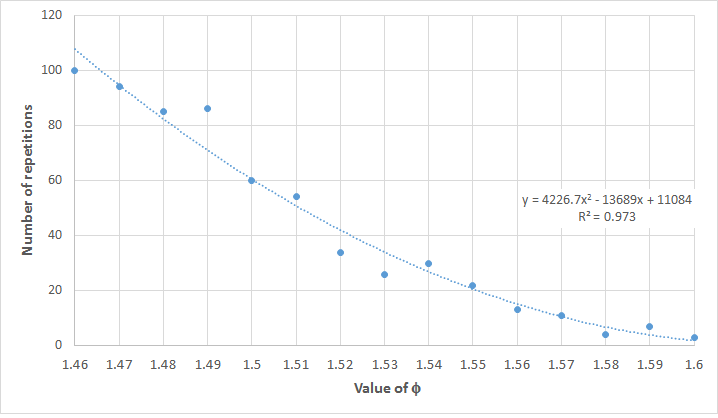}
	\caption{Number of repetitions that converged in terms of $\phi$}
	\label{fig:phivsconverg}
\end{figure}

To better understand the changes between $1.4 \leq \phi \leq 1.6$, we made another experiment with fine grained values for $\phi = \{1.4, 1.41, 1.42, \ldots 1.59, 1.6\}$. 
The results are shown in Table \ref{tab:influNegFeedback2} (right). The simulation converged in all the runs when $\phi \leq 1.46$. For higher $\phi$ values, the number of converged runs reduces following a cuadratic relation ($y=4226.7x^2 - 13689x+11084 \:\:$,$R^2 = 0.973$) (see Fig. \ref{fig:phivsconverg}). Please note that these changes in $\phi$ values imply just a modification of $\gamma$ at the fourth decimal place.

\begin{table}
	\centering
	\footnotesize
\begin{tabular}{@{}l@{}cc@{}r}	
		\begin{tabular}{cc@{}rr}
			$\phi$ & $\gamma$ & $runs$ &I2C \\
			\hline	
0.5	& 0.025	& 100 &52 \\
0.6	& 0.030	& 100 &58 \\
0.7	& 0.035	& 100 &65 \\
0.8	& 0.040	& 100 &73 \\
0.9	& 0.045	& 100 &86 \\
1.0	& 0.050	& 100 &100 \\
1.1	& 0.055	& 100 &127 \\
1.2	& 0.060	& 100 &162 \\
1.3	& 0.065	& 100 &238 \\
\hline
1.4	& 0.070	& 100 &544 \\
1.5	& 0.075	&  60 & 2205 \\
1.6	& 0.080	&  3  &4557 \\
\hline
1.7	& 0.085	&  0  & - \\
1.8	& 0.090	&  0  & - \\
1.9	& 0.095	&  0  & - \\
2.0	& 0.100	&  0  & - \\
			\hline
	\end{tabular} & & &
	\begin{tabular}{ccrr}
	$\phi$ & $\gamma$ & &I2C \\
	\hline	
	1.40 & 0.0700 & 100 & 544 \\
	1.41 & 0.0705 & 100 & 604 \\
	1.42 & 0.0710 & 100  & 722 \\
	1.43 & 0.0715 & 100  & 910 \\
	1.44 & 0.0720 & 100  & 1091 \\
	1.45 & 0.0725 & 100  & 1292 \\ 
	1.46 & 0.0730 & 100  & 1378 \\
	1.47 & 0.0735 & 94   & 1554 \\
	1.48 & 0.0740 & 85   & 2043 \\
	1.49 & 0.0745 & 86   & 2065 \\	
	1.50 & 0.0750	&  60 & 2205 \\
1.51	& 0.0755 & 54 & 2494 \\
1.52	& 0.0760 & 34 & 2324 \\
1.53	& 0.0765 & 26 & 1999 \\
1.54	& 0.0770 & 30 & 2537 \\
1.55	& 0.0775 &  22 & 2206 \\
1.56	& 0.0780 & 13 & 2171 \\
1.57	& 0.0785 & 11 & 3347 \\
1.58	& 0.0790 &  4 & 1028 \\
1.59	& 0.0795 &  7 & 2481 \\
1.60	& 0.080	&  3  &  4557 \\
\end{tabular}\\
\end{tabular}

	\caption{Influence of $\phi$ in the average time to convergence. Values for $I2C$ are rounded. \label{tab:influNegFeedback2}}
\end{table}

These results raise another question: \textit{when a simulation does not converge, which is the average level of trust reached?}

\begin{figure}
	\centering
	\includegraphics[clip=true,width=0.78\linewidth]{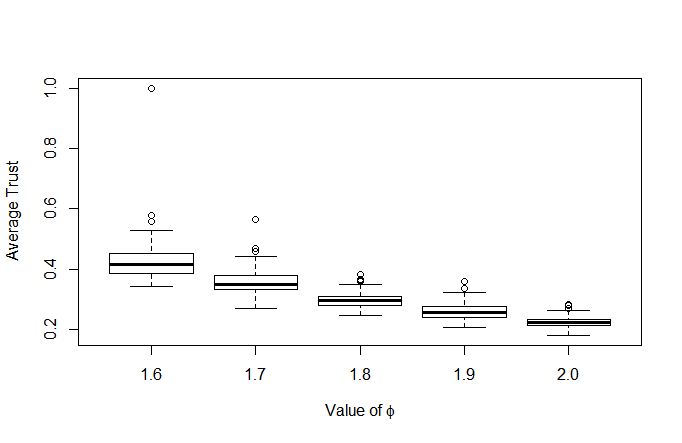}
	\caption{Average trust after 5000 iterations for different values of $\phi$ (Recall $\gamma = \phi \times \beta$)}
	\label{fig:boxplotphi}
\end{figure}

Figure \ref{fig:boxplotphi} shows boxplots corresponding to the average trust values achieved for $\phi \geq 1.6$. It is clear that as the negative feedback increases (the users are more intolerant to recommendation errors), it becomes harder to the average trust to increase. In fact, such value never gets higher than 0.4. 
Recall that when $\phi = 2$, then a recommendation error has two times more impact in the trust than a good recommendation. The plot shows that in this case, the average trust is almost always below than $0.3$ which in turn means that 7 out of 10 recommendations (70\%) are rejected by the users.


\section{Conclusions \label{conclus}}

We focused in an idealized recommender tool that issues binary recommendations over using or not a resource with limited capacity.

We proposed a simple model to study both: 
1) the evolution of trust and 2) the impact of users attitude on the trust dynamics.

We focused on two research questions for which, the main conclusions are outlined.

\noindent
\textit{ 1) Is there any $t^*$ which makes the individual confidences $\alpha_i^{t^*}, \: \forall \: i \in [1 \ldots N]$ converge?. In such a case, the convergence value is the same for all the users?.}

The experiments confirmed that the answer to this question is YES. 
Using neutral users (the same positive and negative learning factors $\beta = \gamma  = 0.05$), all the simulations ended with all the users having $\alpha_i = 1$. In other words, at some point in time, all the users accept the recommendation. 
It was also observed that the value $t^*$ (the time to convergence) follows a power law relation with the number of users.  

From the point of view of the resource usage, this is very important: if the users accept the recommendation, then the recommender can properly balance the attendance to bar. Moreover, the recommendation problem can be transformed onto an assignment problem and then a more ``fair'' approach for recommendations can be implemented (instead of a random one).

\noindent
\textit{ 2) Does the user attitude (the relation between $\gamma$ and $\beta$) has any impact in the time to convergence?}.

The answer is YES. The relation between both parameters has a very strong impact in the time (or number of iterations) to converge. 
Recall that each time the recommender produced a good recommendation, the user's trust is increased in $\beta$ units, while it is decreased by $\gamma$ if the recommendation was bad. 
When $\beta > \gamma$, users are tolerant to recommendation errors. A good recommendation weights more than an error. Under this configuration, the simulation always converged.
As the difference $\gamma - \beta$ became bigger, the number of simulations that converged reduced following a quadratic relation. This is related with the fact that the average trust on the recommendation stayed in low values (below 0.4).

Another important observation is how sensitive is this simple model with respect to small variations in $\gamma$. With $\gamma = 0.07$, all the simulations converged, while using $\gamma = 0.085$ none of them did.

This ``sensitivity to initial conditions'' is a very well known situation in the complex systems field \cite{Mitchell2009}.

If using this simple model, such variations are observed, then one should be very careful when analyzing more complex ones, as very small variations may lead to very big changes in the system behavior.

Overall, we consider this model and the results obtained as first step towards understanding the impact of trust dynamics in  recommendation tools for resources with limited capacity.

\section*{Acknowledgments}
Research supported in part by project TIN2017-86647-P (Spanish
Ministry of Economy and Competitiveness, includes FEDER funds from the European Union).


\end{document}